\documentclass[aps,prb,twocolumn,showpacs]{revtex4}
\usepackage{graphicx}
\usepackage{amsmath}

\begin{document}

\title{Broken Time Translation Symmetry as a model for Quantum State Reduction}

\author{Jasper van Wezel}
  \affiliation{T.C.M. Group, Cavendish Laboratory, University of Cambridge, Cambridge CB3 0HE, UK}

\begin{abstract}
The symmetries that govern the laws of nature can be spontaneously broken, enabling the occurrence of ordered states. Crystals arise from the breaking of translation symmetry, magnets from broken spin rotation symmetry and massive particles break a phase rotation symmetry. Time translation symmetry can be spontaneously broken in exactly the same way. The order associated with this form of spontaneous symmetry breaking is characterised by the emergence of quantum state reduction: systems which spontaneously break time translation symmetry act as ideal measurement machines.
In this review the breaking of time translation symmetry is first compared to that of other symmetries such as spatial translations and rotations. It is then discussed how broken time translation symmetry gives rise to the process of quantum state reduction and how it generates a pointer basis, Born's rule, etc. After a comparison between this model and alternative approaches to the problem of quantum state reduction, the experimental implications and possible tests of broken time translation symmetry in realistic experimental settings are discussed.
\end{abstract}


\maketitle

\section{Introduction}
The phyical laws of nature typically possess a great amount of symmetry. We expect Newton's laws for example to give an adequate description of the motion of a pendulum, irrespective of its precise position on earth or the time of day. Imposing such symmetry constraints on theorems in physics has direct implications for the processes that can be described by them. The translational and temporal invariance of Newton's laws for example are responsible for the conservation of momentum and energy in classical physics. In quantum mechanics the role played by symmetry is even stronger. Having a translationally invariant Hamiltonian does not just forbid processes that don't conserve momentum. It also means that all eigenstates of that Hamiltonian are simultaneously eigenstates of the (total) momentum operator. In principle any physical object described by such a Hamiltonian should therefore always be in a translationally invariant state. Clearly this situation is not realized in our everyday world. Even though the Hamiltonian governing the description of objects like tables and chairs must be translationally invariant due to the homogeniety of empty space, the objects themselves can occur in a state that breaks the translational symmetry. The understanding of the mechanism of spontaneous symmetry breaking, explaining how symmetry-broken states can result from the symmetric laws of nature, is one of the highlights of modern quantum physics.\cite{Anderson1,Anderson2} It was originally formulated in the context of magnetism in solid state theory, but is in fact a general phenomenon that is also central to many of the ideas in other fields of physics, including elementary particle physics and cosmology.\cite{SSB} 

Almost any conceivable form of symmetry is spontaneously broken somewhere in nature, from broken translational symmetry in crystals or broken phase rotation symmetry in superconductors and massive elementary particles to broken supersymmetry resulting in the distinction between bosons and fermions. The only symmetry in quantum mechanics that is not often thought of as being subject to spontaneous breakdown is its unitary time translation symmetry. The mechanism of spontaneous symmetry breaking is traditionally formulated in an equilibrium description, thereby pre-empting any possibility of finding a state that does not obey the unitary symmetry. We have recently shown that this is not a necessary constraint, and that it is possible to give a dynamical description of spontaneous symmetry breaking in quantum mechanics which does indeed allow even the time translation symmetry to break down.\cite{SUB1,SUB2} The time evolution resulting from the spontaneous breakdown of unitarity turns out to be surprisingly familiar: it reproduces precisely the quantum state reduction process observed whenever we try to measure a quantum state with an effectively classical measuring apparatus. 
The aim of this article is to review how certain objects can spontaneously break unitary time translation symmetry, to explain why this implies that these objects can induce quantum state reduction, and to compare the predictions of this model of quantum state reduction whith some of the other models found in the literature.

The paper is organized as follows. We first give a brief overview of the workings of spontaneous symmetry breaking in quantum mechanics. The central concepts in this description are the order parameter field, the singular nature of the thermodynamic limit and the so-called `thin' spectrum. We illustrate these notions using the elementary example of a harmonic crystal which breaks translational symmetry. We then turn to the special case of unitary time translation symmetry, and show that this symmetry too can be broken spontaneously. The description closely follows that of the standard case, and again the roles of the main players (the order parameter field, the singular limits and the thin spectrum) are clarified by applying them to the harmonic crystal. It is also shown how gravity may provide the required order parameter field in this case, due to the inherent conflict between the principles of general covariance and unitarity. The fourth section presents the application of the theory as a model for quantum state reduction. It discusses how the timescales of the non-unitary dynamics can give rise to a distinction between microscopic and macroscopic objects, and points out that this automatically leads to the emergence of a pointer basis and Born's rule, without reference to an environment. After a short discussion of the main principles underlying some other models for quantum state reduction, we then discuss how the predictions of these models can be distinguished from spontaneously broken unitarity using the experimental setting of a mesoscopic object in a state of spatial superposition. We conclude with a summary and outlook to future experiments. 

\section{Spontaneous symmetry breaking}
Both in classical and quantum mechanics it is often possible to find a stable (ground) state of a system which does not respect the symmetry of the physical laws that govern it. The way in which these symmetry-broken states are stablilized in the quantum case is subtly different from their counterpart in the classical case. 

\begin{figure}[t]
      \begin{center}
      \includegraphics[width=0.07\textwidth]{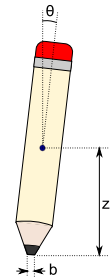}
      \end{center}
      \caption{A nearly balanced pencil. The limits of making the pencil infinitely sharp ($b \to 0$) and perfectly balanced ($\theta \to 0$) do not commute, so that even the smallest defelction will tip over a sharp enough pencil.}
      \label{fig1}
\end{figure}

The generic example of spontaneous symmetry breaking in classical physics is to consider a pencil balanced on its tip. The upright position of the pencil is a metastable state, and the ground state of this classical system is a configuration in which the pencil lies flat on the table. In which direction the pencil points does not matter: all directions form equivalent, degenerate groundstates due to the rotational symmetry of the setup. The pencil, being a classical object, cannot fall towards all directions at the same time. If it leaves its metastable balanced state it will have to pick out one particular direction to fall towards and thus break the rotational symmetry. The broken symmetry can be parameterized by introducing an {\em order parameter} such as the three-dimensional position $(x,y,z)$ of the centre of mass of the pencil. To see how one can go from an unordered, balanced state with $z \neq 0$ and $x=y=0$ to an ordered, symmetry-broken state with $z=0$ but $x$ or $y \neq 0$, consider a pencil that is not perfectly sharp, and not perfectly balanced (as in figure \ref{fig1}). If the flat base of the pencil $b$ is wide enough and the angle of defelction $\theta$ is small enough, the pencil will always remain upright. Only if $\theta$ becomes so great that it tips the centre of mass of the pencil over the end of its base will the pencil drop. If we now take the limit of an infinitely well-balanced, infinitely sharp pencil we find that this is a {\em singular limit}: the fate of the pencil (parametrized by the height of its centre of mass $z$) depends on the order in which we take the limits.
\begin{align}
\lim_{b \to 0} ~ \lim_{\theta \to 0} ~ z &\neq 0 \nonumber \\
\lim_{\theta \to 0} ~ \lim_{b \to 0} ~ z &= 0.
\label{limits}
\end{align}
If the pencil were perfectly balanced, it would always stay upright. But if an infinitely sharp pencil is even only infinitesimally far away from being perfectly centered it will always fall down. We say that in that case the pencil can {\em spontaneously} break the rotational symmetry because any disturbance, no matter how small, will tip it over.

\subsection{The harmonic crystal}
The classical description of spontaneous symmetry breaking is also applicable to some special cases in quantum physics. The Hamiltonians describing for example a $\varphi^4$ field theory or a ferromagnet have many degenerate groundstates. In these cases picking out a single groundstate which breaks the symmetry of the Hamiltonian is very similar to picking out a single orientation for the pencil to fall towards. This, however, is not the generic situation for symmetric quantum systems. In general, a Hamiltonian with a continuous symmetry will have single non-degenerate ground state that obeys the same symmetry. Examples include all types of antiferromagnets, superconductors, Bose-Einstein condensates, crystals, and so on. For those systems, the symmetry broken state is not a ground state of the Hamiltonian. In fact, it usually is not even an eigenstate. The quantum version of the mechanism of spontaneous symmetry breaking will thus have to explain both how a single symmetry-broken state is favoured over all others, and how a state that is not an eigenstate of the Hamiltonian can nonetheless be realized and be stable. To illustrate how this can be done, consider the textbook example of a harmonic crystal, with the Hamiltonian
\begin{align}
H=\sum_{j} \frac{{\bf p}^2_j}{2 m} + \frac{\kappa}{2} \sum_{j}  \left( {\bf x}_j - {\bf x}_{j+1} \right)^2,
\label{eq:Xtal}
\end{align}
where $j$ labels all $N$ atoms in the lattice, which have mass $m$, momentum ${\bf p}_j$ and position ${\bf x}_j$. We consider here only a one-dimensional chain of atoms, but all of the following can be straightforwardly generalized to higher dimensions as well. The harmonic potential between neighboring atoms is parameterized by $\kappa$; it turns out that the results on spontaneous symmetry breaking that follow are equally valid for anharmonic potentials.

In the standard treatment of the harmonic oscillator, one begins by introducing new coordinates, which are the displacements of atoms from their equilibrium positions. After a Fourier transformation the eigenstates of this Hamiltonian are then easily found. This method has the disadvantage that it does not address the motion of the crystal as a whole, but instead only focusses on the internal degrees of freedom (i.e. the phonons). Because the breaking of translational symmetry requires the crystal as a whole to localize at a single position in space, we need to keep track of the external as well as the internal coordinates. This can be done by introducing bosonic operators from the very beginning and diagonalizing the quadratic part of the Hamiltonian by performing a Bogoliubov transformation at the end. This procedure brings to the fore the so called {\em thin spectrum} in a natural manner, and can be easily adapted to the descriptions of antiferromagnets, superconductors and Bose-Einstein condensates.\cite{PRB,ours,ours25,%ours3,ours4,ours5,
Birol}

The momentum and position operators are expressed in terms of bosonic operators as
\begin{align}
{p}_j = i C \sqrt{\frac{\hbar}{2}} (b^{\dagger}_{j} - b^{\phantom\dagger}_j) ;~~{x_j} = \frac{1}{C}\sqrt {\frac{\hbar}{2}} (b^{\dagger}_{j} + b^{\phantom\dagger}_j), \nonumber 
\end{align}
so that the commutation relation  $[{x_j}, {p_{j'}}] = i \hbar \delta_{j,j'}$ is fulfilled. We choose $C^2 = \sqrt{2m \kappa}$ so that the Hamiltonian after a Fourier transformation reduces to
\begin{align} 
H = \hbar \sqrt{\frac{\kappa}{2m}} \sum_k \left[ A_k b^{\dagger}_{k} b^{\phantom\dagger}_k + \frac{B_k}{2} (b^{\dagger}_k b^{\dagger}_{-k} + b^{\phantom\dagger}_k b^{\phantom\dagger}_{-k}) +1 \right], \nonumber 
\end{align}
where $A_k =2 - \cos \left( ka \right)$, $B_{k} = - \cos \left( ka \right)$ and $a$ is the lattice constant. This form of the Hamiltonian can be diagonalized using a Bogoliubov transformation by introducing the transformed boson operators $\beta^{\phantom\dagger}_k=\cosh(u_k) b^{\phantom\dagger}_{-k} + \sinh(u_k) b^{\dagger}_k$, and choosing $u_k$ such that all terms other than the number operator disappear. The result seems to coincide with the textbook Hamiltonian which we would have obtained by following the conventional route:
\begin{align}
H_{k \neq 0} = 2 \hbar {\sqrt \frac{\kappa}{m}} \sum_{k \neq 0} \sin |ka/2| \left[  \beta^{\dagger}_{k} \beta^{\phantom\dagger}_{k} +\frac{1}{2} \right].
\label{knot0}
\end{align}
However, the Bogoliubov transformation has the advantage that it automatically singles out the collective part of the crystal dynamics. When $k \to 0$ both of the parameters  $\sinh(u_k)$ and $\cosh(u_k)$ in the Bogoliubov transformation diverge.\cite{PRB} We should therefore treat the bosonic terms with $k=0$ seperately from  the rest. This zero wavenumber part of the Hamiltonian simply describes the kinetic energy of the crystal as a whole, while the finite wavenumber operators in the Hamiltonian of equation \eqref{knot0} describe the internal dynamics of the crystal in terms of phonons. 

As a function of the original operators, the $k=0$ part can be written as
\begin{align}
H_{{\bf k}=0}= \frac{p^2_{\text{tot}}}{2 N m} + \text{constant},
\label{thin}
\end{align}
where ${\bf p}_{\text{tot}} \equiv \sum_j {\bf p}_j=\sqrt{N} {\bf p}_{{\bf k}=0}$ is the total momentum of the entire system, or equivalently, its center of mass momentum.
It can be straightforwardly checked that the collective Hamiltonian of equation \eqref{thin} commutes with the total Hamiltonian of equation \eqref{eq:Xtal}. The eigenstates of the crystal can thus be labelled by seperate quantum numbers for the centre of mass momentum and the internal phonon excitations. The single, non-degenerate ground state of the harmonic crystal has no phonons in it, and has zero total momentum. It is thus completely delocalized, in accordance with the translational symmetry of its Hamiltonian. The symmetry-broken state, which picks out a definite locus in space, is not an eigenstate of the Hamiltonian, and it can only be formed by a linear combination of different total momentum states. If $N$ is large, these excitations of the crystal which change its centre of mass momentum, but leave the internal state of the phonons untouched, are very low in energy. In fact, the $k=0$ part of the spectrum is called the {\em thin} spectrum because it contains so few states of such low energy that its contribution to the free energy in the limit $N \to \infty$ (the thermodynamic limt) completely disappears.\cite{PRB} In turn, this implies that
these thin spectrum states do not contribute to any thermodynamically measurable quantities such as for instance the specific heat of the crystal. Their effect on the properties of the crystal is increasingly subtle, but its existence can nonetheless have profound consequences, leading for example to decoherence in solid state qubits.\cite{ours,ours25}%,ours3,ours4,ours5}

\subsection{Breaking the symmetry}
The fact that the thin spectrum states all become equal in energy as the system size grows to infinity, makes that limit a singular limit. As in the case of classical symmetry breaking, it is this singularity which enables the system to realize an ordered, symmetry broken state. To see that the formation of the ordered state happens spontaneously, consider the collective part of the Hamiltonian with a very small symmetry breaking field, or {\em order parameter field} added to it
\begin{align}
H'_{{\bf k}=0}= \frac{p^2_{\text{tot}}}{2 N m} + \frac{1}{2} N m \omega^2 {x}_{\text{tot}}^2.
\label{Hsb}
\end{align}
Here ${\bf x}_{\text{tot}}$ is the centre of mass position of the crystal as a whole. Notice that as in the case of the pencil, the disturbance away from the perfectly symmetric system introduced by the order parameter field may not actually exist. It is only introduced here as a mathematical tool to clarify the singularity of the thermodynamic limit, and we will accordingly send $\omega$ to zero at the end of the calculation. The collective Hamiltonian in equation \eqref{Hsb} is just a harmonic oscillator, and its ground state wavefunction is well known to be
\begin{align}
\psi_0 (x_{\text{tot}}) =\left( \frac{N m \omega}{\pi\hbar} \right)^{1/4} e^{-\frac{N m \omega}{2 \hbar} x_{\text{tot}}^2}. \nonumber
\end{align}
This function describes a wavepacket in real space consisting of wavefunctions that differ only in their total momentum. In other words, the order parameter field has introduced a coupling between the thin spectrum states of the symmetric Hamiltonian. We can now again study the fate of this localized, symmetry broken wavefunction under the action of two non-commuting limits:
\begin{align}
\lim_{N \to \infty} ~ \lim_{\omega \to 0} ~ \psi_0 (x_{\text{tot}}) &= \text{const} \nonumber \\
\lim_{\omega \to 0} ~ \lim_{N \to \infty} ~ \psi_0 (x_{\text{tot}}) &= \delta_{{\bf x}_{\text{tot}},0}.
\label{limits2}
\end{align}
The perfectly symmetric Hamiltonian has a perfectly symmetric ground state regardless of how many particles make up the crystal. For an infinitely large crystal however, even an infinitesimally small perturbation of the symmetry is enough to completely localize the crystal in a single position. In that limit we thus again say that the crystal can spontaneously break the symmetry of its underlying Hamiltonian. Notice that the infinitely small order parameter field not only suffices to pick out one point in space above all others, but also to combine the total momentum states which are the eigenstates of the symmetric Hamiltonian into a stable symmetry-broken wavefunction. The reason that only an infitesimal field is necessary to do this, lies in the special properties of the thin spectrum. In the thermodynamic limit all thin spectrum states become degenerate with the ground state, and in that limit it thus costs no energy to form a symmetry breaking wave packet out of them.

Real crystals of course are not infinitely large. Neither can real pencils be made infinitely sharp. What the singularity of the thermodynamic limit means for these real systems is that as long as we consider large enough crystals or sharp enough pencils, almost {\em any} deviation from perfect symmetry will be immedeately reflected by the system. Any such departure from the perfect case will be amplified by a very large factor indeed (either $N$ or $1/b$). Although the symmetry in a finite harmonic crystal described by equation \eqref{eq:Xtal} can strictly speaking not be broken {\em spontaneously}, it can be argued that there will always be some interaction with some other object somewhere in the universe which effectively looks like equation \eqref{Hsb} and which is strong enough to localize the $10^{23}$ particles in a typical real system, just like there is always some deviation from the perfect setup that prevents you from actually balancing a sharp pencil on its tip.

\section{Spontaneously broken unitarity}
Just like the homogeniety and isotropy of empty space enforce any quantum Hamiltonian in it to be translationally and rotationally invariant, the homogeniety and isotropy of time impose their own time-translation and time-inversion symmetries. The equivalence of all directions in time is hard-wired into the formalism of quantum mechanics by the {\em unitarity} of its time evolution, which is in turn guaranteed by the Hermiticity of the Hamiltonian. Despite this stringent constraint, it is conceivable that there could exist situations in which the time dependent wavefunction of a quantum system does not respect the unitarity imposed by its governing Hamiltonian, just like a crystal can exist in a state that defies translational symmetry, and just like a superconductor can ignore the phase symmetry of its governing laws. To find systems that are susceptible to a spontaneous breakdown of unitarity, and the time-dependent states that result from it, we need the same ingredients as in the standard description of spontaneous symmetry breaking: a singular themodynamic limit, an order parameter field, and a thin spectrum.
\begin{figure*}[t]
      \begin{center}
      \includegraphics[width=0.95\textwidth]{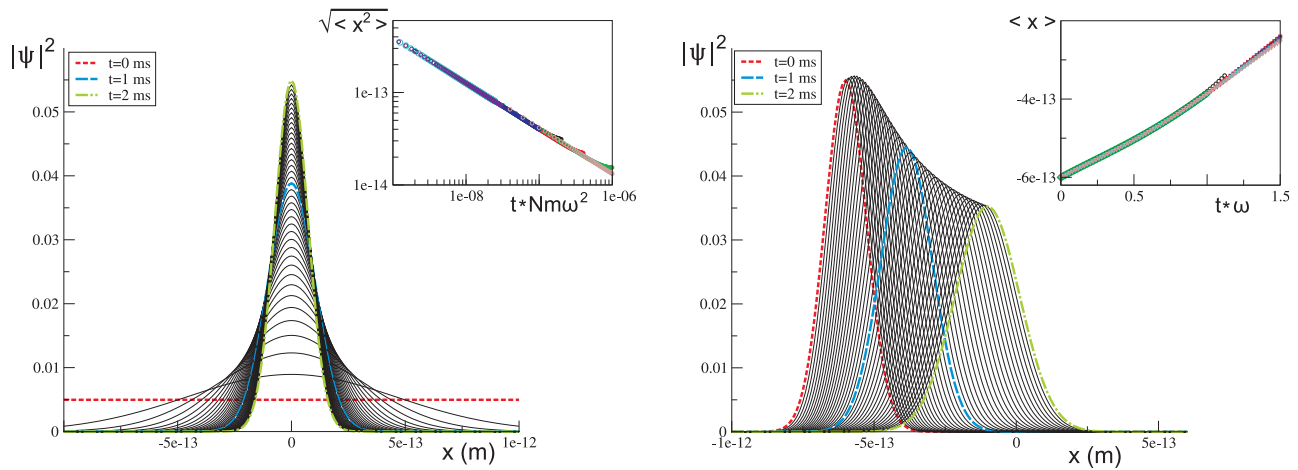}
      \end{center}
      \caption{The dynamics of the crystal wavefunction under the influence of a non-unitary field. {\bf Left}: the evolution of equation \eqref{UB}, starting from a perfectly homogeneous (${\bf p}_{\text{tot}}=0$) initial state, and resulting in a localized state at the origin. The parameters used in this figure are $N m = 1.5 \cdot 10^{-9}$ kg and $\omega=0.1$ kHz. The inset shows the spread in position as a function of $t N m \omega^2$ for several simulations of the same process, with parameters varying between $1.0 \cdot 10^{-11} < N m < 1.0 \cdot 10^{-9}$ kg and $ 0.1 < \omega < 1.0$ kHz. It indicates that the time scale over which the reduction from an unordered, delocalized state to an ordered, localized state takes place, is proportional to $1 / (N m \omega^2)$. The final spread in each individual case is determined by the competition between the two terms in equation \eqref{UB}, and is given by $\langle x^2 \rangle= \hbar \sqrt{1 / N m} / \sqrt{N m \omega^2}  =\hbar / (N m \omega)$. {\bf Right}: the evolution starting from an already localized state. The final state will again be a localized state at the origin, but the translation of the initial state towards the origin requires both terms in equation \eqref{UB}, resulting in a timescale for this process proportional to $1 / \omega$. The parameters used in this figure are $N m = 1.0 \cdot 10^{-11}$ kg and $\omega=1.0$ kHz. The inset shows the average position as a function of $t \omega$ for several simulations of the same process, with parameters varying between $1.0 \cdot 10^{-11} < N m < 1.0 \cdot 10^{-10}$ kg and $ 0.1 < \omega < 1.0$ kHz.}
      \label{fig2}
\end{figure*}

To see how these come together to break the unitarity of quantum time evolution, consider again the example of the harmonic crystal. As shown in the previous section, the collective properties of the crystal as a whole are given by the $k=0$ part of the symmetric Hamiltonian:
\begin{align}
H_{{\bf k}=0}= \frac{p^2_{\text{tot}}}{2 N m},
\label{thin2}
\end{align}
which we called the {\em thin spectrum}, because of its neglicible contribution to the total free energy, and because all of its eigenstates become degenerate in the limit of infinte system size. If we ignore the internal (phonon) dynamics of the crystal, the time evolution of its wavefunction will be given by applying the operator $U(t)=\exp(-i H_{{\bf k}=0} t/\hbar)$. The time evolution will be unitary because $H_{{\bf k}=0}$ is a Hermitian operator. To study the spontaneous breakdown of unitarity, we do the same as in the case of any other symmetry: we add an {\em order parameter field} to the symmetric Hamiltonian of equation \eqref{thin2} which renders the time evolution generated by it non-unitary and we consider the fate of the wavefuncion in the {\em thermodynamic limit} as the strength of the order parameter field is sent back to zero. The unitarity breaking field must couple to the order parameter of the crystal, because only by using the amplification factor $N$ can a field of infinitesimal strength have any effect in the thermodynamic limit. The simplest form of non-unitary time evolution that we can consider is therefore given by
\begin{align}
U'(t)= \exp \left( -i \frac{t}{\hbar} \left[ \frac{p^2_{\text{tot}}}{2 N m} - i \frac{1}{2} N m \omega^2 {x}_{\text{tot}}^2 \right] \right),
\label{UB}
\end{align}
in close analogy to the equilibrium description of broken translational symmetry in equation \eqref{Hsb}.
The fact that the order paremter field only couples to the collective thin spectrum states and not to the internal phonon degrees of freedom ensures that the time evolution described by this equation does not violate conservation of energy in the thermodynamic limit. In that limit all total momentum states become degenerate, and any dynamics involving those states will have a disappearing effect on the total energy. 

\subsection{The time scales of non-unitary dynamics}
The ground state in the presence of the unitarity breaking field will no longer be a static state. Instead we need to consider the time-dependent ground state obtained by applying $U'(t)$ to the ground state of the crystal at some initial time $t=0$. In fact, there are two possible choices for the initial ground state of the crystal from which to start. We can either begin with the exact, symmetric ground state of the collective Hamiltonian, or we could take a symmetry-broken ground state localized at some position in space. Because the factor $\exp(- N m \omega^2 {x}_{\text{tot}}^2 / 2)$ in $U'(t)$ exponentially suppresses all components of the wavefuntion, except for the one localized at $x=0$, it is clear that the wavefunction of the crystal will eventually be reduced to a symmetry broken state centred around ${\bf x}_{\text{tot}}=0$, independent of the intial state. The timescale over which the transformation to the final state takes place however, does depend on the initial configuration of the wavefunction. Because of the opposing influences of the two terms in equation \eqref{UB}, an analytic expression for the dynamics starting from a general initial state is not straightforwardly obtained. The numerical solutions shown in figure \ref{fig2} however, clearly show the difference in timescales for the dynamics of the two initial conditions. 

Starting from a purely symmetric ${\bf p}_{\text{tot}}=0$ state, all positions have equal weight initially and the final state is approached with a half time proportional to $\hbar/(N m \omega^2)$, set entirely by the coupling of the unitarity breaking field to the order parameter. On the other hand, if the initial wavefunction is already localized (at some point $x \neq 0$), there is only a vanishing weight of the wavefunction at the amplified position, and the kinetic energy term in the Hamiltonian is required to spread out the initial wavepacket before the non-unitary dynamics can take place. In that case the approach to the state localized at $x=0$ happens over a timescale proportional to $\hbar / \sqrt{ (1 / N m) (N m \omega^2) } = \hbar / \omega$. The factor $N m$ drops out, and in the limit of vanishing non-unitarity the dynamics is dictated wholy by the Hermitian Hamiltonian of equation \eqref{thin2}. A localized, symmetry broken state is thus stable against non-unitary influences for any number of particles. The delocalized state however, does give rise to a singular thermodynamic limit. The fate of that time-dependent wavefunction depends on the order in which we take the limits of infinite system size and vanishing non-unitarity. We find
\begin{align}
\lim_{N \to \infty} ~ \lim_{\omega \to 0} ~ \psi_0 (x_{\text{tot}},t > 0) &= \text{const} \nonumber \\
\lim_{\omega \to 0} ~ \lim_{N \to \infty} ~ \psi_0 (x_{\text{tot}},t > 0) &= \delta_{{\bf x}_{\text{tot}},0},
\label{limits3}
\end{align}
where $\psi_0 (x_{\text{tot}},t=0) = \text{const}$. From these equations, we conclude that if nature is truely, perfectly symmetric in time, the quantum dynamics of any initial state is unitary for all system sizes. However, an infinitely large crystal in a delocalized initial state will be sensitive to even an infinitesimally small departure from unitarity, and the resulting dynamics will instantaneously reduce a spread-out wavepacket to a localized state. A truely infinitely large crystal can thus spontaneously break the unitarity of its time evolution and localize its wavefunction at a given point in space. Once that has happened, its location is fixed for all time because neither kinetic energy nor an infinitesimal non-unitary field has any influence on the localized state.

\subsection{The order parameter field}
As observed before, real systems are not infinitely large. Strictly speaking, real crystals can therefore not {\em spontaneously} escape the unitarity of quantum time evolution. The meaning of the non-commmuting limits of equation \eqref{limits3} for finite-sized crystals is that as long as we consider large enough crystals, almost {\em any} deviation from unitarity will be immedeately reflected by the system. As in the equilibrium case, any such departure from the perfectly symmetric situation will be amplified by the factor $N$. Real crystals can thus be expected to display non-unitary dynamics only if it can be argued that there will be some field somewhere in the universe whose influence on the crystal effectively looks like equation \eqref{UB}. 

Such influences of course are forbidden by construction in the quantum theory of closed systems, because the Hamiltonian (being an observable) is required from the onset to be purely Hermitian. Such a requirement for unitarity however, is not necessary in all realms of physical law. Einstein's general theory of realtivity for example, is not a unitary theory. In fact, one of the problems in unifying quantum theory with general relativity can be argued to be the incompatibility between the defining properties of general relativity (i.e. its general covariance) and the unitarity of quantum mechanics.\cite{penrose} Building on this observation, it is possible to construct an explicit derivation of how equation \eqref{UB} may arise from the interplay between gravity and quantum mechanics, as will be shown in detail in the next section. 

But even without committing ourselves to any particular source for the order parameter field, it is clear that unitarity as such is not a fundamental prerequisite for all laws of physics. It is therefore not unreasonable to expect that a very small non-Hermitian order parameter field may exist as a consequence of some physical process which is not normally described in terms of quantum mechanics. The singularity of the thermodynamic limit in equation \eqref{limits3} then tells us that no matter how small that order parameter field may be, it will have a noticable effect on the dynamics of a large enough crystal because its influence will be amplified by a factor proportional to the number of particles in the crystal.

\subsection{Gravity's influence on quantum mechanics}
One particular way in which a non-unitary time evolution could arise, is as a consequence of the interplay between general relativity and quantum mechanics. The physical principle that lies at the heart of general relativity is general covariance (or rather, diffeomorphism invariance). That is, the idea that local changes in the choice of coordinate system do not affect the form of physical laws. This principle immediately implies that it is in general impossible to define a globally applicable coordinate frame. It has been argued that the absence of such a global structure makes the principle of general covariance incompatible with the presence of a unitary time evlution, as required by quantum mechanics.\cite{penrose,tjerk} To completely overcome this incompatibility, one would need a full-fledged theory of quantum gravity. 

In the absence of such a theory, it may still be possible to see the first effects that gravity has on quantum mechanics in certain situations by treating them as small perturbations to Schr\"odinger's equation.\cite{tjerk} Let's assume that the inherent non-unitarity of general covariance is the most important characteristic of its conflict with quantum mechanics, and leave other possible ingredients (such as non-linearity) to higher order terms, so that to first order, the perturbed Schr\"odinger equation can be written as
\begin{align}
i\hbar \frac{d}{d t} \psi(\vec{r},t)=\left[ H-iX \right] \psi(\vec{r},t),
\label{iX}
\end{align}
where $H$ is the usual quantum Hamiltonian and $X$ is a Hermitian operator. The fact that the time evolution generated in this way does not conserve energy (as measured by $H$) agrees with the lack of a locally conserved energy concept in a non-static configuration of general relativity. Of course globally energy should be a conserved quantitiy, and we will have to choose $X$ such that it takes account of that restriction. 

\begin{figure}[t]
      \begin{center}
      \includegraphics[width=0.70\columnwidth]{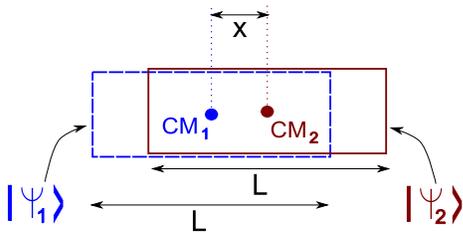}
      \end{center}
      \caption{The massive superposition $|\psi\rangle=\sqrt{1/2}(|\psi_1\rangle+|\psi_2\rangle)$ of a block of width $L$ and mass $M$ over a small distance $x$.}
      \label{fig3}
\end{figure}

To get an idea of what the size of the correction $X$ should be, we need a measure of the incompatibility between general covariance and unitarity. Such a measure has been introduced for the special case of a massive object superposed over two distinct spatial locations.\cite{penrose,philmag,diosi1,diosi2} As an example, consider a block of mass $M$ which is evenly superposed over a distance $x$ small compared to its length $L$ (as in figure \ref{fig3}). For this case, a good measure for the extent to which it is impossible to treat the superposition of spacetimes in a generally covariant manner turns out to be\cite{tjerk,philmag}
\begin{align}
\Delta = G \frac{M^2}{2 L^3} x^2, \nonumber
\end{align}
where $G$ is the gravitational constant, and $\Delta$ has units of energy. The form of this expression allows a straightforward generalization to the case of a generic superposition consisting of any number of components carrying arbitrary weights in the wavefunction. By interpreting $x$ as the quantum mechanical operator which measures the position of the block's centre of mass, the expression can be applied to any wavefunction with an overall centre of mass at $x=0$. The (semiclassical) measure for the ill-definedness of a covariant treatment of a given arbitrary wavefunction of the block in that case reduces to the expectation value of the quantum operator $\Delta$. We can thus identify $\Delta$ with the first order perturbation in equation \eqref{iX}.\cite{tjerk} Once again ignoring the internal properties of the block and considering only its collective dynamics, the perturbed Schr\"odinger equation is then given by
\begin{align}
i\hbar \frac{d}{d t} \psi(\vec{r},t)=\left[\frac{p^2}{2 M} -i \frac{1}{2} G \frac{M^2}{L^3} x^2 \right] \psi(\vec{r},t),
\label{UB2}
\end{align}
which preceisely reproduces equation \eqref{UB} if we take $\omega^2 = G \rho$, with the mass density $\rho \propto M / L^3$.

This expression for the first order effect of gravity on quantum mechanics however, still has one unsatisfactory aspect. As we argued before, the defining characteristic of the theory of general relativity is its general covariance, which implies the impossibility of defining a globally applicable coordinate frame within a given spacetime. Considering the superposition of spacetimes associated with the different positions of the block, the absence of a gobally defined coordinate system in either of the components makes it impossible in general to identify a particular point in one spacetime with any one point in the other. The best we can possibly do is make an approximate identification of different regions by referring to their local spacetime structure. At best there is thus a many-to-many identification of spacetime points in different components of the wavefunction, rather than a one-to-one correspondence, due to the requirement of general covariance.\cite{penrose,tjerk} The implication of this for the expression in equation \eqref{UB2} is that the operator $x$, which measures the position of the block's center of mass within one component of the wavefunction with respect to the overall average center of mass at $x=0$, becomes inherently ill-defined. To model this ill-definedness of the position operator, we introduce a small {\em stochastic variable} $\xi(t)$ which randomly fluctuates in time, and replace $x$ with $x-\xi$.

Notice that both general relativity and quantum mechanics are in fact fully deterministic theories. The introduction of a random variable here should be seen only as a poor man's approach towards a resolution between the two theories: although superpositions and random variables may not actually feature in the exact reality of quantum gravity, we assume that if we insist on the possibility of effectively describing the state of the system using the concept of superpositions --even though they are ill-defined notions in General Relativity-- then we also need to take into account an effectively random correction to the notion of position. We are thus led to a minimal correction to the unitary Schr\"odinger equation due to the influence of general covariance, which for the case of the massive block in our example reads
\begin{align}
i\hbar \frac{d}{d t} \psi(\vec{r},t)=\left[\frac{p^2}{2 M} -i \frac{1}{2} G \frac{M^2}{L^3} \left(x-\xi(t) \right)^2 \right] \psi(\vec{r},t).
\label{UB3}
\end{align}

The phenomenological treatment of the first order effects of gravity in equation \eqref{UB3} can be seen as a specific realization of the order parameter field of equation \eqref{UB}. It shows that massive crystals may be able to spontaneously break the unitarity of their quantum time evolution due to the influence of gravity. At the moment, it is unclear if there exists a similar treatment for the order parameter fields associated with different ordered states such as antiferromagnets or superconductors. Perhaps they can indirectly couple to gravity,\cite{tjerk} or perhaps there is another realm of physics that could somehow provide their specific order parameter fields. For now, we will leave this question open and focus solely on the case of the harmonic crystal, for which general covariance provides the justification for the introduction of a non-unitary term.

\section{Spontaneously broken unitarity as a model for quantum state reduction}
In the previous section we showed that the singularity of the thermodynamic limit implies that any unitarity breaking order parameter, even if it is only infinitesimally small, will have a noticable effect on the dynamics of a large enough crystal. Reversing the argument, we can also posit that the dynamics of quantum objects may be used as a {\em detector} for non-Hermitian order parameter fields. After all, if any such field does exist, it will inevitably render the dynamics of a large enough crystal non-unitary. There is ample experimental confirmation that the unitary theory of quantum mechanics correctly predicts the dynamics of objects ranging from elementary particles to C$_{60}$ molecules and even coherent supercurrents of more than $10^{6}$ electrons.\cite{Youngs,Zeilinger,Mooij} At the same time, it is also well known that truely macroscopic objects do not seem to obey Schr\"odinger's equation. Taken at face value, the fact that we can use a macroscopic apparatus to project a microscopic quantum state onto a given set of basis states suggests that the apparatus undergoes some form of non-unitary dynamics.\cite{vNeumann} This observation that the phenomenon of quantum state reduction does not immedeately fit into the standard framework of the quantum theory lies at the basis of the many different `interpretations' of quantum mechanics. Despite the succes of some of these interpretations in explaining certain aspects of the problem, we still lack conclusive evidence in favour of any one particular interpretation, and no general concensus on the subject has been reached. In the light of the continuing mystery surrounding the problem of quantum state reduction, one may wonder if the sensitivity to non-unitary influences experienced by large enough crystals could help to understand the observed behaviour of macroscopic objects. The influence of a non-Hermitian order parameter field which is too small to have any noticeble effect on microscopic particles could become observable when multiplied by the number of particles in a typical macroscopic quantum measurement machine, and perhaps explain its observed non-unitary dynamics.

Apart from providing a dynamical description of the quantum state reduction process, there are three basic requirements that any potential resolution of this problem needs to fulfill. Any theory of quantum state reduction must first of all provide an explanation of why only macroscopic objects seem to be affected by it, and what exactly defines an object as being macroscopic. Secondly, it should determine which states are the possible outcomes of a measurement process (the so-called Pointer basis \cite{Zurek}), and how this basis of states is selected over all other possibilities. Finally, it needs to give rise to Born's rule. That is, the probability for a particular result of a measurement to be realized should equal the squared amplitude of the corresponding component in the initial wavefunction of the quantum state to be measured.\cite{Born}

\subsection{The dynamics of quantum state reduction}
We have already seen in the previous section that spontaneous unitarity breaking can account for some difference between the dynamics of microscopic objects and that of macroscopic ones. The time it takes for a delocalized state to spontaneously localize due the presence of an order parameter field is proportional to $\hbar / (N m \omega^2)$. If $\omega^2$ is of the correct order of magnitude, this timescale could be unmeasurably long for microscopic objects while being unmeasurable short for macroscopic objects.\cite{SUB1,SUB2} If we assume that the order parameter field results from the incompatibility of general covariance and unitarity, as we argued before, then the combination of $\omega^2 = G \rho$ and $M$ is indeed of the correct order of magnitude to ensure that elementary particles, molecules and even supercurrents should be considered microscopic, while tables, chairs and pointers should be treated as macroscopic objects. This division however only applies to the dynamics starting from a fully delocalized state. To account for the seperation between microscopic and macroscopic objects during a general quantum state reduction process, we need to consider a more general initial state of the crystal. 

Let's assume that we want to measure some property of a microscopic particle by coupling it to the position of a macroscopic crystal. We should then arrange things in such a way that the final position of the crystal (at $x=x_0 \pm a$) tells us the inital state ($|\phi=\pm 1\rangle$) of the microscopic particle:
\begin{align}
| \phi=+1 \rangle \otimes | x=x_0 \rangle &\to | \phi=+1 \rangle \otimes | x=x_0+a \rangle \nonumber \\
| \phi=-1 \rangle \otimes | x=x_0 \rangle &\to | \phi=-1 \rangle \otimes | x=x_0-a \rangle. \nonumber
\end{align}
\begin{figure}[t]
      \begin{center}
      \includegraphics[width=0.95\columnwidth]{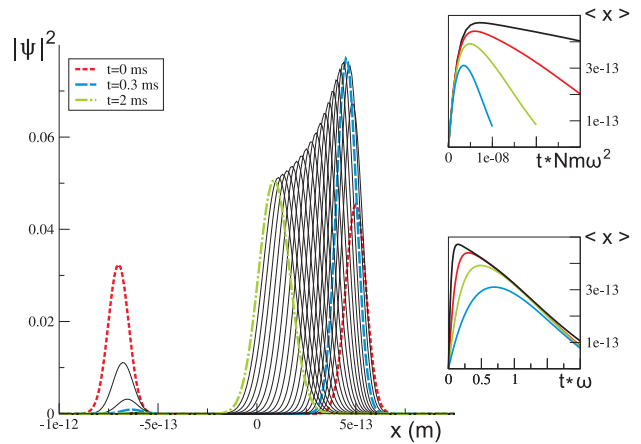}
      \end{center}
      \caption{The dynamics of the crystal wavefunction under the influence of a non-unitary field, starting from an initial superposition of two localized states. The final state, localized at the origin, is reached via two consecutive processes. First the superposition is reduced to just a single localized wavefunction within a timescale proportional to $1/(N m \omega^2 )$, analogous to the process in the left of figure \ref{fig2}. Then the single localized state is translated towards the origin within a time scale proportional to $1/ \omega$, as in the right of figure \ref{fig2}. The parameter values used in this figure are $Nm = 2.0 \cdot 10^{-11}$~kg and $\omega = 1.0$~kHz. The insets show the average position as a function of $t N m \omega^2$ (top) and $t \omega$ (bottom), for several simulations of the same process, with parameters varying between $0.5 \cdot 10^{-11} < N m < 5.0 \cdot 10^{-11}$ kg while $ \omega = 1.0$ kHz.}
      \label{fig4}
\end{figure}
In the absence of an order parameter field, the unitarity of time evolution in quantum mechanics then implies that measuring a superposed microscopic state with the same setup must result in a superposed macroscopic state.
\begin{align}
 \left( \ \alpha \ | +1 \rangle + \beta \ | -1 \rangle \right) \otimes | x_0 \rangle & \to  \nonumber \\
 \alpha \ | +1 \rangle \otimes | x_0+a \rangle & + \beta \ | -1 \rangle \otimes | x_0-a \rangle. \nonumber
\end{align}
In the presence of a small order parameter field, the states corresponding to single localized positions of the crystal are stable and take a time proportianal to $\hbar / \omega$ to feel the effect of the non-unitarity. The final state in the last measurement however cannot be a stable state. This superposition of the crystal over two distinct locations is not the completely delocalized state studied before, but the effect of the non-unitary field is similar. After all, the non-unitary part of equation \eqref{UB} will suppress any component of the wavefunction which is not at $x=0$. Whichever of the two components at $x=x_0\pm a$ lies furthest from the central point will be suppressed most. In fact, the suppression of the furthest component will be exponentially stronger than that of the nearest. We might therefore expect one of the components of the wavefunction to dominate after a typical timescale which is again set only by the non-unitary field and its coupling to the order parameter. The numerical solution of the dynamics of the superposed state in figure \ref{fig4} shows that this is indeed the case. An initial state consisting of a superposition over any number of different localized states reduces to just a single of its initial components under the time evolution of equation \eqref{UB} within a time proportional to $\hbar / (N m \omega^2)$. After that, the single localized state further evolves into a wavefunction localized at the centre of the order parameter field in a timescale proportional to $\hbar / \omega$.\cite{SUB1,SUB2,tjerk}

\subsection{quantum measurement}
These timescales invite the following interpretation of the measurement process in terms of the spontaneous breakdown of unitarity. Initially a microscopic object may be in any quantum state without being affected by the order parameter field because the timescale proportional to $\hbar/ (N m \omega^2)$ is unmeasurably long. The macroscopic measurement machine will initially be in a state of spontaneously broken symmetry, and it too will be unaffected by the non-unitarity, because its single, ordered state will only conform to the non-unitary potential over the timescale proportional $\hbar  / \omega$, which is too long to be observed. The measurement procedure then consists of a quantum mechanical interaction which forces the state of the macroscopic measurement machine to reflect the initial state of the microscopic object. If the initial state of that object was a superposition, then the resulting state of the macroscopic object will also be a superposed state. This coupling process is assumed to happen very rapidly, so that the order parameter field will not noticably affect the dynamics until after the macroscopic wavefunction has been put into a superposition. Once that has happened though, the macroscopic superposed state is found to be unstable, and its evolution will be dominated by the unitarity breaking order parameter field which suppresses all but one component of the wavefunction. The one component surviving after a timescale proportional to $\hbar/ (N m \omega^2)$ is a localized, ordered state which corresponds to a single component of the initial wavefunction of the microscopic object.

This description of measurement makes a distinction between microscopic objects, which are so small that they cannot spontaneously break the unitary symmetry of time evolution (or any other continuous symmetry), and macroscopic objects, which are so large that they are almost infinitely sensitive to the order parameter field and almost instantaneously break unitarity whenever they are forced into a superposition of ordered states. In between these two extreme cases there may be mesoscopic objects, which do respond to the presence of a non-Hermitian field, but take a finite (and possibly measurable) time to do so. If we assume the order parameter field to originate from gravity, it can be straightforwardly shown that these mesoscopic states fall precisely in the unexplored experimental gap between objects which are already proven to act purely quantum mechanically (such as C$_{60}$ molecules and supercurrents) and objects which are expected to always behave classically (such as tables and pointers).\cite{penrose} 

The second requirement on the description of quantum state reduction, that of determining a basis of possible final outcomes of a measurement process, is also automatically fulfilled by the invocation of spontaneous unitarity breaking. The only macroscopic wavefunctions that are stable against the influence of the order parameter field are the ordered, localized states of the crystal. These are precisely the symmetry-broken states (or generalized coherent states) that we customarily refer to as classical states. They form a literal pointer basis in the sense that they describe all possible single positions that the pointer could take without allowing any superpositions of them. A quantum measurement machine in this description is thus necessarily a classical object displaying a spontaneously broken symmetry. The outcomes of measurements can be recorded in a pointer basis only if the coupling to the microscopic quantum state is such that it affects the order parameter associated with the broken symmetry of the measurement machine. As it turns out, all known quantum measurement machines are of this type, mainly because the human senses are only capable of directly registering classical (order parameter) properties of macroscopic machines. Without coupling to some classical order paramter, we would not be able to record the outcome of any measurement.

\subsection{Born's rule}
\begin{figure}[t]
      \begin{center}
      \includegraphics[width=0.95\columnwidth]{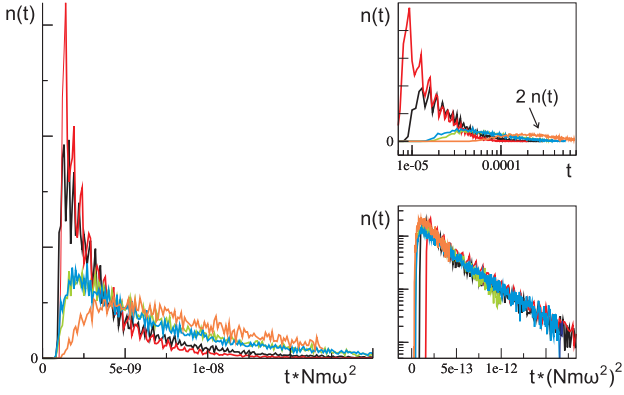}
      \end{center}
      \caption{The distribution of times after which the crystal dynamics in the presence of a fluctuating field yields a localized wavefunction. The normalized number density $n(t)$ (in arbitrary units) indicates the number of simulations out of $50000$ in which a single component first dominates the wavefunction at time $t$. The data is plotted as a function of $t$ in the upper inset. The same data is shown in the main figure, but as a function of $t N m \omega^2$, indicating that the onset of the distribution is proportional to $1/( N m \omega^2)$. The lower inset displays the same data again, but plotted against $t (N m \omega^2)^2$, showing that the width of the distribution is proportional to $1/( N m \omega^2)^2$ . The different lines represent sets of simulations with parameter values ranging between $0.5 < \omega < 1.0$~kHz and $0.5 \cdot 10^{-11} < N m < 1.5 \cdot 10^{-11}$~kg.}
      \label{fig5}
\end{figure}

As it stands, the third requirement of reproducing Born's rule has not been fulfilled. If the non-unitary field is a given, static field, it will always favour the component of a superposed wavefunction that is closest to its centre. Even if the location of that centre changes from one run of the experiment to another, this could never lead to a distribution of experimental outcomes that is correlated to the weight of individual components in the initial wavefunction.\cite{SUB1,SUB2,philmag2} However, we saw in the previous section that the conflict between general covariance and quantum mechanics can give rise to a non-unitary perturbation of Schr\"odinger's equation that is best modelled by the introduction of a time dependent stochastic variable, as in equation \eqref{UB3}. The time dependent nature of this variable implies that the component of the initial wavefunction which is favoured by the non-unitarity at one instant in time, may be suppressed the next. Its random nature on the other hand makes it impossible to predict deterministically which component will win out in which run of the experiment. The best we can do is make a statistical prediction about the probability that a certain component will dominate the wavefunction after a time proportional to $\hbar / (N m \omega^2)$ (that this is indeed the correct time scale for the selection of a single component, even in the presence of a randomly fluctuating field, can be tested numerically, as shown in figure \ref{fig5}). 

It turns out to be possible to make these statistical predictions without solving the detailed dynamics of equation \eqref{UB3} by introducing an auxiliary, `external' object in an entangled state with the crystal.\cite{SUB1,SUB2,Zurek2} During the time evolution described by the perturbed Schr\"odinger equation, the external object does not interact with the crystal, nor with the microscopic state to be measured. In fact, it is completely absent from the perturbed Hamiltonian, and may not actually exist. It is introduced here only as a mathematical tool to assist in the discussion of the crystal dynamics. It is sufficient for the following discussion that such an entangled, external object could in principle exist, even if it is perhaps never actually present.

Consider the initial wavefunction
\begin{align}
| \psi \rangle = \alpha | e_1 \rangle \otimes |x=a> + \beta |e_2 \rangle \otimes |x=b>, \nonumber
\end{align}
where $a \neq b$ and $|x=a,b \rangle$ denotes the state of the crystal localized at $x=a,b$, and $|e_{1,2}\rangle$ is the state of the external object. The time evolution of this initial state governed by equation \eqref{UB3} does not depend on any of the properties of the external object. It acts solely on the crystal's wavefunction and, as we showed in the previous section, reduces the superposed state to just one of its components within the (short) timescale proportional to $\hbar /  (N m \omega^2)$. The external object being unaffected by these dynamics, the total final state will thus be either $| e_1 \rangle \otimes |x=a>$ or $|e_2 \rangle \otimes |x=b>$. The probabilities $P_a(\psi)$ and $P_b(\psi)$ for ending up in the former or the latter state respectively, cannot depend on any property of the external object. They must also be independent of the non-unitary field, since its centre is determined by the fluctuating random variable $\xi$ which is guaranteed by symmetry not to favour any one particular outcome. The only quantity that $P_{a,b}(\psi)$ can possibly depend on, is the form of the initial state, parameterized by the quantities $\alpha$ and $\beta$. Because of this, the initial state
\begin{align}
| \varphi \rangle = \alpha | e_3 \rangle \otimes |x=a> + \beta |e_4 \rangle \otimes |x=b> \nonumber
\end{align}
must also give rise to the probabilities $P_a(\varphi)=P_a(\psi)$ and $P_b(\varphi)=P_b(\psi)$, independent of the external states. If we consider the special case with $| e_3 \rangle=e^{i \theta}| e_1 \rangle$ and $| e_4 \rangle=| e_2 \rangle$, this shows that the probabilities cannot depend on the phases of $\alpha$ and $\beta$.

Next, consider the initial states
\begin{align}
| \psi \rangle &= \alpha | e_1 \rangle \otimes |x=a> + \beta |e_2 \rangle \otimes |x=b>, \nonumber \\
| \chi \rangle &= \beta | e_1 \rangle \otimes |x=a> + \alpha |e_2 \rangle \otimes |x=b>. \nonumber
\end{align}
It is immediately clear that $P_a(\psi)=P_b(\chi)$ for any choice of $\alpha$ and $\beta$. In the special case $|\alpha|=|\beta|$, we also know that $P_a(\psi)=P_a(\chi)$, so that it is clear that in that case we must have $P_a(\psi)=P_b(\psi)$, and thus the perhaps trivial looking result that equal sizes of the initial weights lead to equal probabilities for finding the corresponding final states. This statement can be trivially extended to yield the rule that a set of possible final states with equal weights in the initial wave function leads to equal probability for finding any one of the final states within that set. Continuing that line of thought, consider the initial state
\begin{align}
| \phi \rangle = \alpha |x=a> + \alpha |x=b> +\alpha |x=c> + \text{.. ,} \nonumber
\end{align}
with $a \neq b \neq c$. The combined probability $P_{a \text{ or } b}(\phi)$ must then be equal to $P_{a}(\phi)+P_{b}(\phi)=2P_{c}(\phi)$, which follows directly from the additivity of probabilities and the fact that only a single components of the initial wavefunction can dominate the final state. Extending this result, we find that within a set of possible final states with equal weights in the initial wave function, a subset has a combined probability equal to the relative size of the subset times the total probability of the entire set.

Finally, consider the initial state
\begin{align}
| \psi \rangle = \sqrt{\frac{m}{N}} | e_1 \rangle \otimes |x=a> + \sqrt{\frac{n}{N}} |e_2 \rangle \otimes |x=b>, \nonumber
\end{align}
with $m$, $n$ and $N$ positive integers. The probabilities $P_{a,b}(\psi)$ are again independent of the external states. We are therefore free to write $|e_1\rangle$ and $|e_2\rangle$ in a basis in which they are a sum of states with equal weights:\cite{Zurek2,Zurek3}
\begin{align}
| e_1 \rangle &\equiv \sqrt{\frac{1}{m}} \left[ | E_1 \rangle + | E_2 \rangle + .. + | E_m \rangle \right] , \nonumber \\
| e_2 \rangle &\equiv \sqrt{\frac{1}{n}} \left[ | F_1 \rangle + | F_2 \rangle + .. + | F_n \rangle \right] . \nonumber 
\end{align}
In terms of these states, the expression for the initial wavefunction then becomes
\begin{align}
| \psi \rangle = \sqrt{\frac{1}{N}}\left[ \sum_{i=1}^m | E_i \rangle \otimes |x=a> + \sum_{j=1}^n |F_j \rangle \otimes |x=b> \right]. \nonumber
\end{align}
In this expression all weights are equal, and using the previously found rules we must conclude that $P_a(\psi)=\frac{n}{m} P_b(\psi)$. In the case that the total probability for finding any outcome at all is $1$, this result precisely corresponds to Born's rule: the probability for finding a specific final pointer state is equal to the square of the weight of the corresponding state in the initial wavefunction.\cite{Born} The extension of this result to include also weights which are square roots of non-rational numbers is trivial because the rational numbers are dense on the real line.\cite{Zurek2,Zurek3}

Combining all of these results, the spontaneous breakdown of the unitary symmetry of quantum mechanical time evolution is seen to lead to a natural explanation of all of the phenomena normally associated with the problem of quantum state reduction. It shows why there is a divide between macroscopic and microscopic objects, and gives a definition of where the division occurs. It guarantees that the macroscopic objects can not exist in a (stable) superposition of classical states, while leaving the microscopic objects free to explore the entire Hilbert space associated with the unitary Schr\"odinger equation. It also provides a reason for why only a certain basis of classical states can serve as the final pointer states in a quantum measurement experiment, and establishes which states form that basis. Finally, it correctly predicts the outcome of the quantum state reduction process to be in accordance with Born's rule. Spontaneous unitarity breaking can thus be seen as a model for quantum state reduction, on equal footing with established interpretations of quantum mechanics such as the GRW and CSL models, or decoherence, hidden variable and many world theories. Which of these models gives the most accurate description of reality will have to be decided by experimental evidence focussing on those regions where the theories differ in their predictions.

\section{Comparison to other models of quantum state reduction}
Before discussing the possibility of experimentally distinguishing different models of quantum state reduction, let's first briefly review their main assumptions and predictions. The following will necessarily be incomplete, and will not do justice to the vast amount of physical detail underlying these models, nor to their many individual accomplishments. Many good and thorough reviews and discussions of each of the models exist in the literature.\cite{GRW,CSL,CSL2,GRWCSL,Pearle,Zurek, Zurek2, Zurek3, zeh1, zeh2, caldeira, joos,Adler,deBroglie,bohm,bell,cushing,valentini,Everett, Deutsch, Wallace, Saunders} Here, we only present the essential ideas underlying some specific models for quantum state reduction, and the most prominent features of their predictions, with the aim of finding ways to experimentally differentiate between them.

\subsubsection{The GRW and CSL models}
The GRW (after Ghirardi, Rimini and Weber) and CSL (Continuous Spontaneous Localization) models are very close to the model of spontaneously broken unitarity presented here, in the sense that they hypothesize the existece of a process that falls outside of the realm of applicability of Schr\"odinger's equation which causes the quantum state reduction.\cite{GRW,CSL,CSL2,GRWCSL} In the GRW theory this extra process is assumed to be the spontaneous and instantaneous spatial localization of an elementary particle at particular intervals in time.\cite{GRW} What causes the localization, and which particles it acts on precisely, is not specified. If the frequency with which these localization events occur is low enough, it will take an unmeasurably long time (on average) for any individual particle to undergo such an event. Within an extended object consisting of a macroscopic number of elementary particles on the other hand, it will not take long before at least one of the particles is localized. If we assume that the macroscopic object is rigid in the sense that it describes a (symmetry broken) crystal whose internal wavefunction looks like an ordered array of constituents, but whose overall, external position is still undefined, then the localization of a single particle within the crystal suffices to give the entire wavefunction a definite position in space. GRW thus provides a differentiation between microscopic and macroscopic objects in terms of the typical frequency with which localization events occur. It assumes a basis of pointer states by assuming the perturbation to quantum mechanics to cause events which localize the wavefunction in real space rather than for example in momentum space. It also assumes Born's rule from the outset, by assuming that the probability for a localization event to occur at a particular position is proportional to the squared weight of the wavefunction at that position.\cite{GRWCSL}

The CSL models can be seen as extensions of the GRW model in which the addition to quantum mechanics is no longer a set of instantaneous localization events, but rather a continuous process which constantly acts to gradually localize the individual particles.\cite{CSL,CSL2} In different CSL models this continuous process has been linked to for example the local number density, the local mass density, or a non-linear form of the effective gravitational potential.\cite{GRWCSL,diosi1,diosi2} The average rate of localization is usually a free parameter of the theory that can be chosen to ensure that the localization of microscopic particles will take an unmeasurably long time. In close analogy to the GRW model, macroscopic objects, which are again assumed to posses some amount of rigidity, amplify the localization rate due to their large number of constituents, and are therefore almost instantaneously brought into a single locus. The pointer basis of CSL models is determined by the specific form of the localization process, and is usually taken to be the position basis. Born's rule can spontaneously emerge from the stochastic localization processes described in CSL models due to the competition between possible outcomes, which is precisely analogous to the ``gambler's ruin'' problem in probability theory.\cite{Pearle}

\subsubsection{Decoherence}
An alternative approach to the problems posed by quantum state reduction is based on the phenomenon of decoherence. The basis of decoherence theory is the observation that the exact quantum state of a macroscopic object necessarily involves the description of many microscopic degrees of freedom, most of which are beyond the reach of our current observational skills.\cite{Zurek, Zurek2, Zurek3, zeh1, zeh2, caldeira, joos} If one averages over all the possible values that these unobservable microscopic degrees of freedom can take, the remaining density matrix describing only the observable macroscopic degrees of freedom looks fully classical. It is thus argued that real macroscopic objects can occur in any quantum state, localized or not, but that the macroscopic observables associated with that object seem to take on only classical values if one disregards the microscopic degrees of freedom. Macroscopic objects are then classical `for all practicle purposes' only. Notice that the unobservable (microscopic) traits do not need to be internal degrees of freedom of a macroscopic object. A single elementary particle for example can in the same way behave purely classically by itself if its wavefunction is entangled with enough unobservable particles in its environment. 

The theory of decoherence and the effectively classical behaviour of objects which are strongly entangled with their environments have been experimentally verified many times. In fact, in modern quantum information science, decoherence forms one of the major fundamental stumbling blocks to be overcome in the race towards a working quantum computer. Its application to the problem of quantum state reduction however, is limited by the fact that it only describes the properties of objects when averaged over all possible states of the microscopic or environmental degrees of freedom. In a single experiment, only one specific state of the environment is realised, not an average over all possible states. Decoherence therefore cannot explain the absence of macroscopic superpositions from any single experiment.\cite{Adler} For the average result of many quantum measurements, it does provide all the necessary ingredients: microscopic and macroscopic objects are distinguished by the time it takes them to loose their coherence; the pointer basis is that set of (macroscopic) states which is unaffected by the averaging over environmental degrees of freedom; and Born's rule arises automatically from the averaging procedure.\cite{Zurek2,Zurek3}

\subsubsection{The hidden variable and many worlds interpretations}
Hidden variable theories take the view that Schr\"odinger's equation can be interpreted as describing not just a wavefunction, but also a configuration of particles with definite positions as well as definite velocities. The wavefunction in such a description becomes a `pilot wave' which guides the particle configuration through phase space, much like the force fields of Newtonian mechanics guide classical configurations of particles.\cite{deBroglie,bohm,bell,cushing,valentini} Experiments on quantum mechanical systems are assumed to measure the actual configuration of the particles rather than the state of the wavefunction. The problem of quantum state reduction thus becomes non-existent (the wavefunction is unaffected by measurement), as does the difference between microscopic and macroscopic objects (both consist only of a configuration of point particles). A pointer basis is implicitly acquired via the classicality of the posited particles. To make the predictions of hidden variable theories agree with Born's rule however, an additional assumption is needed. One can either explicitly assume that the intial state of all particles agreed with Born's rule and show that this agreement is then conserved in time;\cite{bohm} or one can use the theory of decoherence to argue that any initial configuration of particles will eventually come to look like it agrees with Born's rule in a course grained description which averages over the microscopic degrees of freedom of the wavefunction.\cite{valentini}

Like hidden variable theories, the many worlds interpretation assumes that the dynamics of all objects, microscopic or macroscopic, can be described by Schr\"odinger's equation. Rather than introducing a particle configuration however, it is argued that the theory can be made to agree with all experimental observations by including the state of the observer in the wavefunction.\cite{Everett} In the course of a typical experiment, the observer and the observed object become entangled. Relative to the state of the observer then, the observed property takes on a definite value in each run of the experiment. All different components of the wavefunction, with all their different states of the observer, and all their different values for the observed quantity are assumend to be equally real (hence `many worlds'), but the observer is only conscious of a single outcome for each macroscopic variable in each component. The existence of a pointer basis is assumed in this description, either by assuming that the human brain itself cannot be in a superposition of different states of mind, or by reference to decoherence processes. Likewise, Born's rule can be attributed either to game theoretical arguments or to the loss of coherence between individual components of the wavefunction.\cite{Everett, Deutsch, Wallace, Saunders}

\subsection{Experimental implications}
Let us now return to the model of spontaneous unitarity breaking, and discuss how its predictions for possible experimental implications differ from those made by the other theories of quantum state reduction. If unitarity is not a fundamental property of nature, and the effect of the existence of non-unitary laws of physics on quantum dynamics can be modelled by the introduction of a non-Hermitian order parameter field such as in equations \eqref{UB} and \eqref{UB3}, then there must be an energy (or mass) scale intermediate between microscopic particles (which for all practicle purposes look purely quantum mechanical) and macroscopic objects (which seem wholly classical). This mesoscopic mass scale, at which the non-unitary dynamics happens on human timescales, must lie beyond the experiments that have been done to date. It should involve massive objects that are large compared to single molecules but small compared to, say, dust particles. In the ideal experimental test of non-unitary dynamics, one would create a superposition of a massive object over two distinct spatial locations, and follow its time evolution. The preparation of the initial state could involve a coupling to a suitably prepared microscopic object, while the monitoring of the dynamics could be done by repeating the experiment many times and projecting the state of the mesoscopic object using a standard (macroscopic) quantum measurement machine after different intervals in each run. Depending on the size of the mesoscopic object, the dynamics should then turn out to be unitary (in the quantum regime), non-unitary (in the unitarity breaking regime), or non-existent (in the classical regime, where the initial state is already a static projection). Some experiments of this type have recently been proposed using a mesoscopic mirror or a magnetized mechanical resonator as the mesoscopic object to be superposed.\cite{tjerk,Bouwmeester}

The difficulty in actually constructing such an experiment lies in the presence of decoherence. Because one cannot monitor the time dependent evolution of the mesoscopic state without disturbing it, we are forced to infer the dynamics from many projections in many different runs of the experiment. Such an ensemble averaged evolution will always look classical if one does not keep track of all the individual degrees of freedom in the environment. Fortunately the decoherence time over which this environment induced process takes place can be calculated using standard methods.\cite{caldeira} Seperating decoherence from the similar effects induced by a non-unitary order parameter field can then be done by ensuring that the decoherence time is longer than the expected timescale of the unitarity breaking dynamics. This typically involves cooling the mescoscopic object to temperatures extremely close to absolute zero.\cite{tjerk,Bouwmeester} Further confirmation about whether the measured dynamics is due to either decoherence or broken unitarity can be gained from exploring the scaling of the quantum state reduction time with the involved mass or geometry of the mesoscopic object. Decoherence effects are usually only weakly dependent on the geometry. If on the other hand, the non-untarity arises from gravity, it will typically depend strongly on the precise geometry of the experimental setup.\cite{philmag}

The non-unitarities of the type predicted by GRW and CSL models can likewise be distinguished by referring to the scaling of the reduction time with parameters of the model and the geometry of the mesoscopic object. GRW predicts that the reduction time will only scale with the number of particles involved, rather than the mass or shape of the object. Different CSL models may predict dependences on number density, mass density, and even geometry, and will thus be more difficult to distinguish from the model of spontaneously broken unitarity. The detailed dynamics of the mesoscopic object however, depend on the precise form of the non-unitary interaction, and thus provides a means of differentiating between the models.

If nature does turn out to be fundamentally unitary, and we have succesfully circumvened the problem of decoherence by cooling the experimental setup to low enough temperatures, it should be possible to create stable superposed states of mesoscopic objects. In that case, there is even no objection in principle to extending the experiment to macroscopic length scales. Given enough time and experimental progress it may then be possible to cool even a chair to sufficiently low temperatures to force its wavefunction into a state of spatial superposition and to do interference expieriments with it. In that case, the apparent absence of macroscopic quantum dynamics from daily experience would be a strong indicator that either a type of hidden variables or a many worlds interpretation of quatum mechanics may be applicable.

\section{Conclusions}
In this paper we have reviewed the possibility of the existence of broken unitarity. We have shown that a description of the spontaneous loss of unitary time translation symmetry in large enough objects can be formulated in close analogy to the standard theory of spontaneous symmetry breaking. The same three main ingredients that are responsible for the breakdown of translational symmetry in crystals, rotational symmetry in magnets, and so on, are also responsible for the loss of unitarity in the time evolution of these objects. These ingredients are the singular nature of the thermodynamic limit, the existence of a `thin' spectrum of extremely low energy states in macroscopic systems, and the presence of an infinitesimally small order parameter field.

The order parameter field responsible for a breakdown of unitarity would have to be a non-Hermitian contribution to the quantum Hamiltonian. We have shown how such a contribution could in principle arise as a first order correction to Schr\"odinger's equation due to the existence of general relativity and the incompatibility between its idea of general covariance and the unitarity of quantum theory. In the presence of a unitarity breaking order parameter field (regardless of its origin), we have shown that the dynamics of a macroscopic, symmetry-broken wavefunction mimics the ideal quantum measurement. Such a wavefunction can no longer appear in a stable superpositon of different symmetry-broken states. If it is forced into such a superposed state by being coupled to a microscopic quantum state, it will quickly reduce to just one of its components, with the probability for ending up in any particular state given by Born's rule. That is, the probability is equal to the square weight of the corresping component in the initial wavefunction of the microscopic object.

We have argued that this description of spontaneously broken unitarity thus fulfills all the requirements for being a model of quantum state reduction. It explains why there is a difference between microscopic and macroscopic objects in terms of the singularity of the thermodynamic limit and the influence of the order parameter field. It gives rise to a pointer basis of states that are stable under the non-unitary time evolution. And it leads to the emergence of Born's rule if macroscopic objects are used as quantum measurement machines.

Comparing the predictions of a spontaneously broken unitarity with those of other models for quantum state reduction, it appears that they can be distinguished if we have access to an experimental setup in which a mesoscopic object can be brought into a state of spatial superposition. If decoherence can be avoided by carefully cooling and isolating the object, it becomes possible to distuinguish the detailed time evolutions predicted by different models of quantum state reduction via their dependence on external parameters such as the shape and mass of the superposed object. Experiments of this type are currently being developed in various places. Some of the most promising ones may be the recent proposals for creating a superposition of a mesoscopic cantilever over different angles of deflection, either by coupling it to light or to a the magnetic field of a suitably prepared spin state.\cite{Bouwmeester, tjerk}

In conclusion, it is suggested in this review that the unitary time translation symmetry of quantum theory may be just another symmetry, which can spontaneously break down if an appropriate order parameter field presents itself. The consequence of losing unitarity in this way would be the emergence of quantum state reduction. Whether or not the non-Hermitian order parameter field actually exists, and whether it is actually responsible for quantum state reduction will have to be decided by future experiments.

%Quantum states which break the unitary symmetry will spontaneously favour one direction of time evolution over the other and thus define an arrow of time.

\section*{Acknowledgements}
\noindent
I am grateful to Jeroen van den Brink, Jan Zaanen, Tjerk Oosterkamp and Wojciech Zurek for many stimulating discussions about this subject, and Homerton College of the University of Cambridge for its financial support.

\end{document}